\documentclass{kapproc} 

\usepackage{procps} 

\usepackage[dvips]{graphicx}

\upperandlowercase

\kluwerbib 

\begin{document}

\articletitle[Dwarf and giant elliptical galaxies]
{On the unification of dwarf and giant elliptical galaxies}

\author{Alister W.\ Graham\altaffilmark{1} and Rafael Guzman\altaffilmark{1}} 
 
\affil{
\altaffilmark{1}
Department of Astronomy, University of Florida, Gainesville, FL 32611, USA.
}

\begin{abstract}
The near orthogonal distributions of dwarf elliptical (dE) and giant
elliptical (E) galaxies in the $\mu_e$--$M$ and $\mu_e$--$\log R_e$
diagrams have been interpreted as evidence for two distinct galaxy 
formation processes.  However, 
%
%
continuous, linear relationships across the alleged dE/E boundary at
$M_B = -18$ mag --- such as the relationships between central
surface brightness ($\mu_0$) and: a) galaxy magnitude ($M$); and b)
light-profile shape ($n$) --- suggest a similar initial formation
mechanism. 
Here we explain how these latter two trends in fact necessitate a different
behavior for dE and E galaxies, exactly as observed, in
diagrams involving $\mu_e$ (and also $<$$\mu$$>$$_e$).
%
A natural consequence is that the location of dEs and Es in 
Fundamental Plane type analyses that use $I_e$, or $<$$I$$>$$_e$, 
will also be different. 
Together with other linear trends across the alleged dE/E boundary,
such as those between luminosity and color, metallicity, 
and velocity dispersion, it appears that the dEs form a 
continuous extension to the E galaxies.  The presence of partially 
depleted cores in luminous ($M_B < -20.5$ mag) Es does however 
signify the action of a different physical process at the 
centers ($< \sim$300 pc) of these galaxies. 
\end{abstract}


The common distinction between a dwarf elliptical (dE) galaxy and an
(ordinary) elliptical (E) galaxy is whether the absolute magnitude is
fainter or brighter than $M_B=-18$ mag respectively ($H_0=50$ km 
s$^{-1}$ Mpc$^{-1}$, Sandage \&
Binggeli 1984).  By the term dE, we additionally mean objects brighter
than -13 $B$-mag; that is, we are not talking about (Local Group)
dwarf spheroidal galaxies, whose range of colors suggest a range of
formation processes (e.g., Conselice 2002). 
%
%
The realization that dE light-profiles could be 
reasonably well described with an exponential function (Faber \& Lin
1983; Binggeli, Sandage \& Tarenghi 1984) and that bright ellipticals
are better fit with de Vaucouleurs' $r^{1/4}$-law 
helped lead to the notion that they are two distinct families of
galaxies (e.g., Wirth \& Gallagher 1984, but see Graham 2002).
One of the seminal papers supporting this view is Kormendy (1985).  By
plotting central surface brightness against luminosity, Kormendy
showed two relations, almost at right angles to each other: one for
the dE galaxies and the other for the luminous elliptical galaxies.
Similar diagrams using $\mu_e$, the surface brightness at the
effective half-light radius $r_e$, or $<$$\mu$$>$$_e$, the average
surface brightness within $r_e$, also show two somewhat perpendicular
relations (e.g., Capaccioli, Caon, \& D'Onofrio 1992).  These
differences are commonly interpreted as evidence for different
formation mechanisms, resulting in the belief that a dichotomy exists
between the dE and E galaxies. 
To understand, and in fact resolve, this apparent dichotomy, 
we must turn to the issue of galaxy structure.

In the past, some authors have restricted the radial extent of galaxy 
light-profiles (excluding inner and outer parts; e.g, Burkert 1993) or
adjusted the sky-background levels (e.g., Tonry et al.\ 1997) in order
to make the $r^{1/4}$ model fit --- such was the ingrained belief in
this classic model.
However, luminosity-dependent deviations from
$r^{1/4}$ profiles had been known for some time
(e.g., Capaccioli 1984, 1987; Michard 1985; Schombert 1986;
Caldwell \& Bothun 1987; Kormendy \& Djorgovski 1989; Binggeli \&
Cameron 1991; James 1991).  
Schombert (1986) recognized the inadequacy of the
$r^{1/4}$ model for describing Es, since it only fits the
middle $21 < \mu_B < 25$ part of bright galaxy profiles.  Kormendy \&
Djorgovski (1989) noted that the best $r^{1/4}$ fits were for
Es with $M_B \sim -21$ mag; brighter and fainter
galaxies having a different logarithmic profile curvature than that of
the $r^{1/4}$ model.  It is these variations in the stellar
distribution which have recently provided the key to understanding the
true nature of the connection between the dE and E galaxies.

\begin{figure}[ht]
\centerline{\includegraphics[height=4.7in,angle=270]{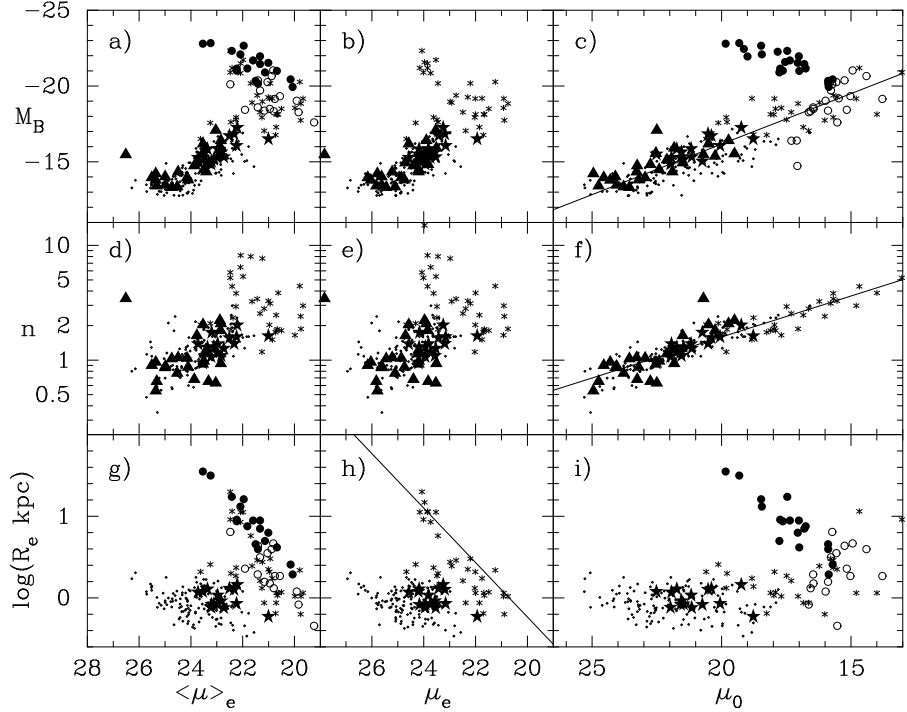}}
\caption{
{\small Mean surface brightness within $r_e$ ($<$$\mu$$>$$_e$), 
surface brightness at $r_e$ ($\mu_e$), and central host galaxy
surface brightness ($\mu_0$) versus host galaxy
magnitude ($M_B$), global profile shape ($n$), and half-light
radius ($r_e$).  Due to
biasing from the magnitude cutoff at $M_B\sim -13$, the line $M_B =
(2/3)\mu_0 - 29.5$ in panel c) has been estimated by eye rather than
by a linear regression routine.  The line $\mu_0=22.8-14\log(n)$ in
panel f) has also been estimated by eye.  The line in panel h) has a
slope of 3 and represents the Kormendy (1977) relation known to fit
the luminous elliptical galaxies which define the panhandle of this
complex distribution (Capaccioli \& Caon 1991; La Barbera et al.\
2002).  Dots represent dE galaxies from Binggeli \& Jerjen (1998),
triangles are dE galaxies from Stiavelli et al.\ (2001), large
stars are dE galaxies from Graham \& Guzman (2003),
asterix are intermediate
to bright E galaxies from Caon et al.\ (1993) and D'Onofrio et al.\
(1994), open circles represent the so-called ``power-law'' E galaxies
from Faber et al.\ (1997), and the filled circles represent the
``core'' E galaxies from these same authors. 
Figure taken from Graham \& Guzman (2003). $H_0=70$ km s$^{-1}$ Mpc$^{-1}$.}
}
\label{fig1}
\end{figure}

Sersic's (1968) $r^{1/n}$ model can encompass both de Vaucouleurs'
$r^{1/4}$ model and the exponential ($n$=1) model --- and a variety of
other profile shapes --- by varying its `shape parameter' $n$. 
Analyzing a sample of 80 early-type galaxies in the Virgo and Fornax
Clusters, Caon, Capaccioli, \& D'Onofrio (1993) and D'Onofrio,
Capaccioli, \& Caon (1994) showed how the elliptical galaxy
light-profile shapes vary systematically with measurements of the 
half-light galactic radii and luminosity obtained 
independently of the fitted Sersic model.
%
%
Additionally, numerous studies have demonstrated that dE galaxy
profiles are \textit{not} universally exponential, but rather are best
fit with a range of Sersic profiles (i.e., $n$ is not always $= 1$;
Davies et al.\ 1988; Cellone et al.\ 1994; Young \& Currie 1994; 
Durrell 1997; Jerjen \& Binggeli 1997;
Binggeli \& Jerjen 1998; Graham \& Guzman 2003).  The resulting 
trend between luminosity and light-profile shape has begun to
erase the dichotomy between dwarf and luminous ellipticals.  
To show how the remaining dichotomies can be eliminated, by explaining
the apparently divergent behavior of dE and E galaxies in certain
structural parameter diagrams, we will use the compilation of 249 dE
and E galaxies presented in Graham \& Guzman (2003) and shown here in
Fig.~\ref{fig1}.

\begin{figure}[ht]
\vskip.2in
\centerline{\includegraphics[height=4.7in,angle=270]{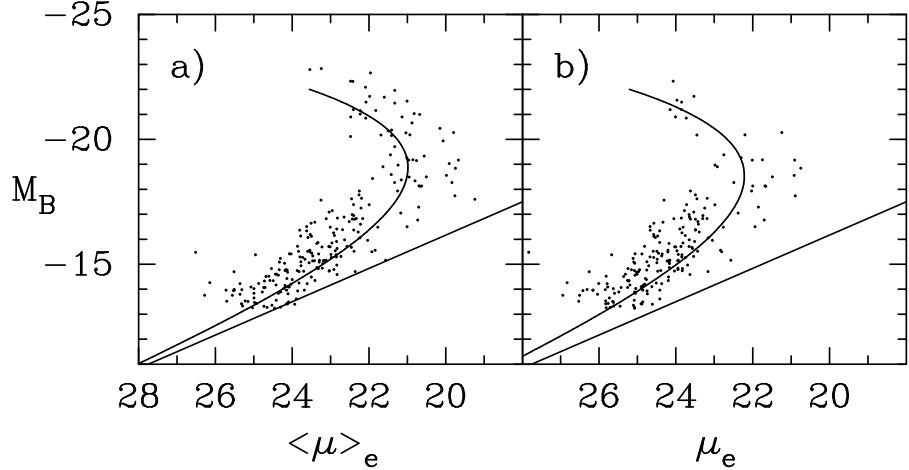}}
\caption{
{\small The curved lines show the predicted behavior of galaxies in
the a) magnitude--mean surface brightness and b) magnitude--effective
surface brightness diagrams.  The straight line is from
Fig.~\ref{fig1}c; the data points are from Fig.~\ref{fig1}a 
and \ref{fig1}b.} 
}
\label{fig2}
\end{figure}

Kormendy's (1985) plot of central surface brightness ($\mu_0$) 
versus magnitude showed a large
discontinuity and gap between the dE and E galaxies.  However, there
was an absence of galaxies with magnitudes around $M_B=-18\pm1$ in
that sample, exactly where one might expect to see the two groups
connect.
If we remove galaxies with $M_B=-18\pm1$ from our plot of 
$\mu_0$ vs.\ $M_B$
(Fig.~\ref{fig1}c), we obtain a figure very much like
Kormendy's (1985) Figure~3.  Thus we can see that part of the
``discontinuity'' had arisen from an incomplete sampling of 
the intermediate-luminosity galaxies which fill in the apparent gap.
Nevertheless, there is still an obvious change in slope in the overall
$\mu_{0}$--$M_{B}$ relation.  This can be explained with the
observation that the most luminous elliptical galaxies possess partially
evacuated ``cores''.  Their observed central surface brightnesses are
thought to be fainter than the original value due to the damage from 
coalescing supermassive black holes following a galaxy merger. 
As stressed in Graham \& Guzman (2003), the brighter galaxies lying
perpendicular to the $\mu_{0}$--$M_{B}$ relation defined by the less
luminous ($M_B > -20.5$) elliptical galaxies are all ``core''
galaxies.  The simplest explanation is that all elliptical galaxies
follow a linear $\mu_{0}$--$M_{B}$ trend, \textit{except} when core
formation modifies the central surface brightness of 
the most luminous ellipticals.  Since core formation is
thought to be a later event, the initial mechanism(s) of dE and E
galaxy formation are likely to be the same. 

Another reason why E galaxies are thought to be different
from dE galaxies is because they don't follow the same $M_B$--$\mu_e$
and $M_B$--$<$$\mu$$>$$_e$ relations.  Why
are these relations
apparently different for the dE and E galaxies?  This, it turns out,
has nothing to do with core formation but is due to the systematic
changes in profile shape with galaxy magnitude.
Even though many of the galaxies in Fig.~\ref{fig1}a--c lack
Sersic $n$ measurements, the relationship between magnitude and $n$
(e.g., Fig.10 in Graham \& Guzman 2003) can be used to determine
a representative value of $n$ for a given $M_B$.  From the Sersic
model, we know that $\mu_e = \mu_0 + 1.086b$ and $<\mu>_e = \mu_e
-2.5\log[e^{b}n\Gamma(2n)/b^{2n}]$, where $b\sim 2n-1/3$ (e.g., 
Graham \& Colless 1997).  From this, the straight line in
Fig.~\ref{fig1}c transforms into the 
\textit{curved} relationships between $M_B$ and $\mu_e$, and 
$M_B$ and $<$$\mu$$>$$_e$ (Fig.~\ref{fig2}).  Thus, we
can reproduce the observed correlations in Fig.~\ref{fig1}a and
\ref{fig1}b.  The different slopes 
for the dE and E galaxy distributions in these diagrams are merely a consequence
of a continuously varying profile shape with galaxy luminosity ---
they do not imply distinctly different galaxy formation processes
for dEs and Es.  
Using $L=2\pi r_e^2$$<$$I$$>$$_e$, it can be shown that 
the same mechanism is also behind the different slopes in the
$<$$\mu$$>$$_e$--$\log r_e$ and $M-\log r_e$ diagrams. 
A natural consequence is that the location of dEs and Es in 
Fundamental Plane type analyses that use $I_e$, or $<$$I$$>$$_e$, 
will also be different.

\begin{acknowledgments}
We are grateful for funding provided by NASA through grants
HST-AR-08750.02-A and HST-AR-09927.01-A administered by the Space
Telescope Science Institute.  A.G.\ is also thankful for NSF funding
administered by the American Astronomical Society's International 
Travel Grant Program.
\end{acknowledgments}

\end{document}